\documentclass[submit]{epsv8}

\usepackage{lineno}
\usepackage{graphicx}
\usepackage{bm}
\usepackage{hyperref}
\usepackage[authoryear]{natbib}
\title{Predictability of a solar flare in May 2024 using observational data-driven MHD simulations}
\author{
Takafumi Kaneko${}^{1,2}$, Niigata University, 8050 Ikarashi 2-no-cho, Nishi-ku, Niigata, 950-2181, Japan,
kaneko@ed.niigata-u.ac.jp
}

\abstract{
We examined the applicability of observational data-driven magnetohydrodynamic (MHD) simulations to flare prediction. The target event was the X1.6 flare that occurred in NOAA AR 13663 at 02:22 UT on 3 May 2024. We employed a velocity-driven model, in which the photospheric velocity field was derived from the time-series magnetograms to use as the boundary input. The simulation showed a rapid increase in both thermal and kinetic energy density around the actual onset time and location of the X1.6 flare. We revealed that the flare was triggered by two-step reconnection, in which the initial tether-cutting reconnection facilitated the subsequent breakout reconnection. We further examined whether the flare could be reproduced when the boundary input was stopped prior to the actual flare onset time, assuming the situation in which the flare must be predicted using the data before it actually occurs. When the photospheric magnetic field was fixed more than 1 hour before the actual flare onset time, the flare was not reproduced in the simulations. In contrast, when the photospheric velocity field at the final observation time was incorporated to infer the subsequent magnetic evolution, the prediction lead time could be extended beyond 1 hour. On the other hand, quantitative prediction of the magnitude of flares remains a subject for future study.
}

\keywords{Solar flares, MHD simulations, Data-driven, Space weather}

\begin{document}

\maketitle

\section{Introduction}

Solar flares are explosive events in the solar atmosphere characterized by rapid and intense enhancements of electromagnetic emission. 
Strong X-ray and high energy particles of the solar flares are hazardous for not only the astronauts in space but also the crews in aircrafts at high-altitude. The Coronal Mass Ejections (CMEs) associated with flares can change near-Earth plasma environment, leading to disruptions of satellite operations.
Since the flares and CMEs have significant impacts on our modern society,
the development of reliable prediction method has been attracting growing interest in both academic research and operational applications.

It is widely accepted that the solar flares and plasma eruptions are driven by rapid change of magnetic field via magnetic reconnection and magnetohydrodynamic (MHD) instabilities \citep{2011LRSP....8....6S,2015SoPh..290.3457S,2017PhPl...24i0501C}. Assessing the magnetic field conditions in the solar atmosphere is essential for understanding the mechanisms and for predicting flares and eruptions. While two-dimensional distribution of the photospheric magnetic fields can be measured directly, it is difficult to observe three-dimensional structure of the coronal magnetic fields above the photosphere. 
Typically, three-dimensional coronal magnetic fields are inferred by numerical modeling based on MHD theory. One standard way to infer coronal magnetic field is the nonlinear force free field (NLFFF) model in which the three-dimensional magnetic fields are extrapolated from the observed photospheric magnetic field at a certain time assuming force-free (mechanical equilibrium) state \citep[e.g.][]{2006SoPh..235..161S,2012ApJ...755...62J,2014ApJ...780..101I}. The other approach is the data-driven method in which time series data of the photospheric magnetic field are incorporated for the bottom boundary input \citep[e.g.][]{2019SoPh..294...41P,2020ApJ...890..103T,2022Innov...300236J}. Since the data-driven method does not assume force-free state, it is expected to have less physical assumption and better temporal consistency.

Considering practical application to flare forecast operations, we have to rely on the observational data obtained before the flare onset. In observational data-driven MHD simulations, even if the boundary input of observed photospheric magnetogram is stopped, the coronal magnetic field within the simulation domain can continue to evolve according to the MHD equations. This post-input evolution may lead to reproduction of a flare in the near future. 

In this study, we examined whether a data-driven simulation can reproduce a flare when the input of the magnetogram is stopped prior to the actual flare onset time. We tested two different types of boundary inputs and investigated the possible lead time for flare prediction within our current data-driven MHD approach.

In Section \ref{sec:event}, we present the observational properties of the flare and the magnetic field of the active region. In Section \ref{sec:method}, we describe the numerical method and setup of case study. Section \ref{sec:result} shows the simulation results. Discussion is provided in Section \ref{sec:discussion}, and the conclusions are summarized in Section \ref{sec:conclusion}.

\section{Target Event}\label{sec:event}
The target event of this study was the X1.6 flare that occurred in  NOAA Active Region (AR) 13663 at 02:24 UT on 3 May 2024.
In May 2024, many X-class flares were observed. The most flare-productive region AR 13664, located in southern hemisphere, produced 15 X-class flares \citep{2025ApJ...979...49H}. AR 13663, located in northern hemisphere, produced 5 X-class flares. The X1.6 flare at 02:24 UT on 3 May 2024 was the first X-class flare from AR 13663 and also the first among all X-class flares in May 2024.   In contrast, no X-class flare occurred in April 2024, although large ARs such as 13643 and 13645 were present and produced M-class flares. Since few major flares had occurred beforehand, predicting the X1.6 flare from AR 13663 was practically more challenging than for the subsequent other flares.

\begin{figure}
    \begin{center}
    \includegraphics[scale=0.6,bb=0 0 614 561]{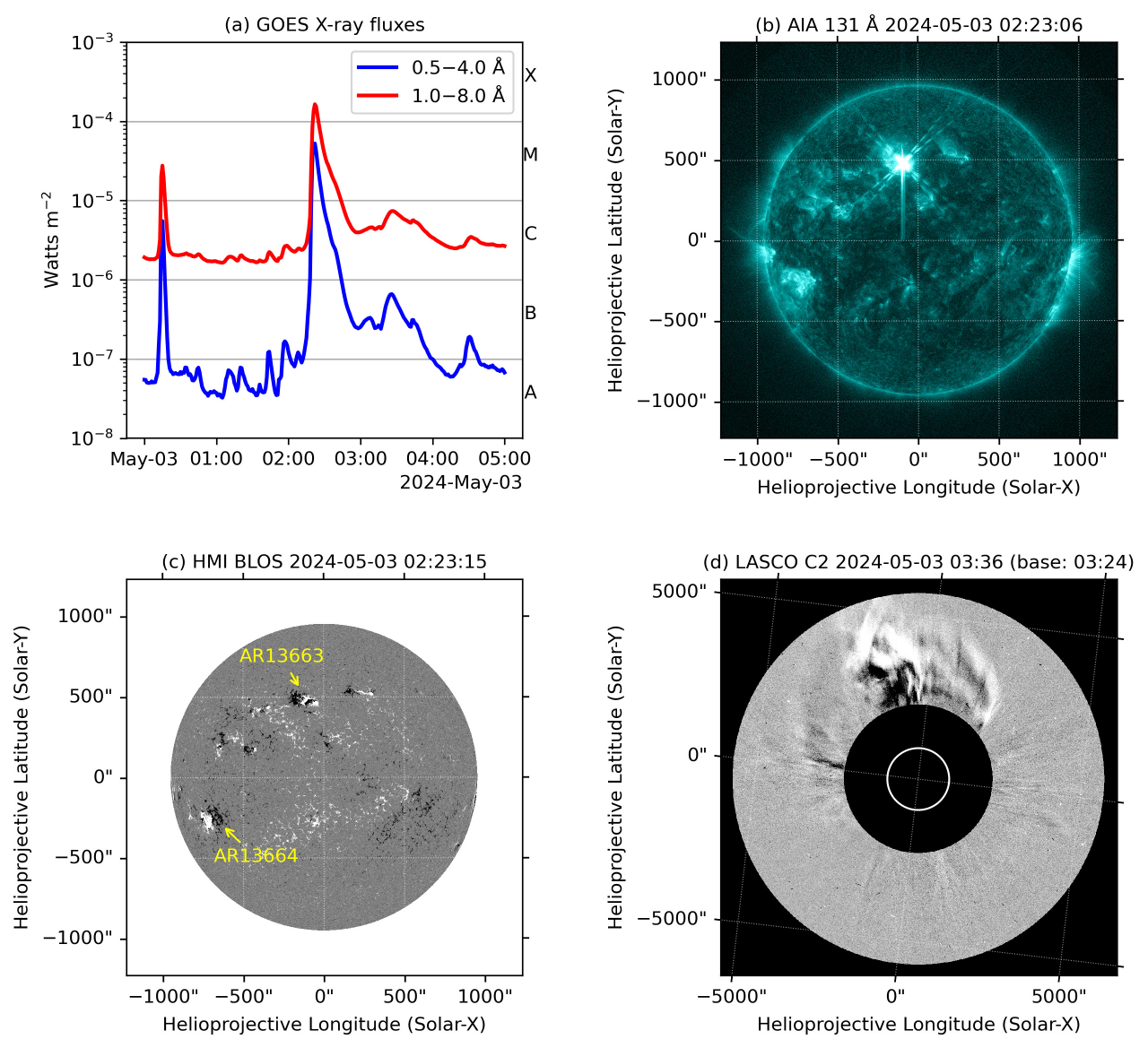}
    \caption{Panel (a): Soft X-ray flux observed by GOES. Panel (b): EUV image observed by SDO/AIA 131 \AA~channel. Panel (c): LOS component of the photospheric magnetic field observed by SDO/HMI. (d) Base-difference image of SOHO/LASCO C2 at 03:36 UT on 3 May. The base image was taken at 03:24 UT.}
    \label{fig:overview}
    \end{center}
\end{figure}

Figure \ref{fig:overview} shows an overview of the target event. Panel (a ) shows soft X-ray flux observed by Geostationary Operational Environmental Satellite (GOES).
The start, peak, and end times of the X1.6 flare were 02:11 UT, 02:22 UT, and 02:27 UT, respectively. 
Panel (b) shows the EUV image at the flare peak time observed by Atmospheric Imaging Assembly \citep[AIA;][]{2012SoPh..275...17L} onboard the Solar Dynamics Observatory \citep[SDO;][]{2012SoPh..275....3P}. 
Panel (c) shows the line-of-sight component of the photospheric magnetic field at the same time as panel (b) observed by Helioseismic and Magnetic Imager \citep[HMI;][]{2012SoPh..275..229S} onboard SDO. We find that the flare occurred in AR 13663 by comparing panels (b) and (c).
Panel (d) shows a CME near the northeast limb observed shortly after the flare by Large Angle and Spectrometric Coronagraph experiment \citep[LASCO;][]{1995SoPh..162..357B} onboard Solar and Heliospheric Observatory \citep[SOHO;][]{1995SoPh..162....1D}. From the location and timing, the CME is assumed to be associated with the X1.6 flare.
Note that the M2.7 flare at 00:15 UT was produced by another active region AR 13664 in the southern hemisphere.

\begin{figure}
    \begin{center}
    \includegraphics[scale=0.8,bb=0 0 488 663]{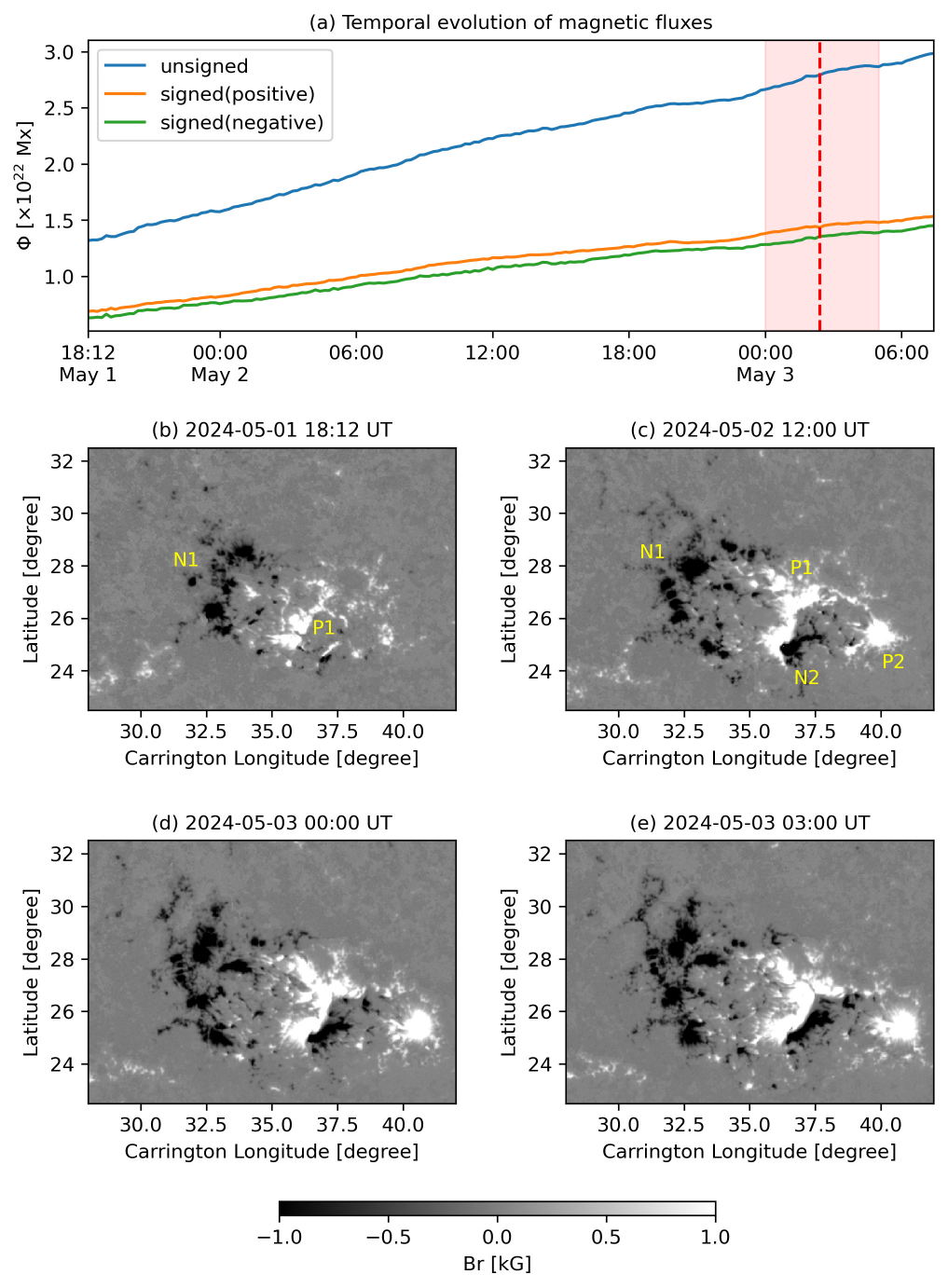}
    \caption{Panel (a) shows the magnetic fluxes of AR 13663 within the field of view shown in panels (b)--(e). The blue, orange, and green lines represent the unsigned, positive, and negative magnetic fluxes, respectively. The vertical red dashed line denotes the peak time of the flare. The red shaded area indicates the time interval during which the data-driven simulation was performed. Panels (b)--(e) represent the radial component of the photospheric magnetic field observed by SDO/HMI.} 
    \label{fig:mag_evol}
    \end{center}
\end{figure}

Figure \ref{fig:mag_evol} shows the evolution of AR 13663 from 1 to 3 May. We obtained the data from the Space-Weather HMI Active Region Patches (SHARP) pipeline \citep{2014SoPh..289.3549B}. Panel (a) shows the temporal evolution of the magnetic fluxes computed using the high-confidence pixels ($\mathrm{conf\_disambig} = 90$). The flare occurred during the flux emergence. Panels (b)--(e) depict the radial component of the photospheric magnetic field observed by HMI from 1 to 3 May. The location of AR 13663 was N25E34, N26E25 and N26E10 on 1, 2, and 3 May, respectively, according to the Solar Region Summary by NOAA.
On 1 May, only a single bipole, P1--N1 in panel (b), was present, and the AR was classified as a $\beta$-spot.
After 18:12 UT on 1 May, another bipole, P2--N2, started to emerge nearby P1--N1, forming a quadrupolar magnetic configuration as shown in panel (c). The spot type was $\beta \gamma$ on 2 May. The flux emergence of P2--N2 continued throughout 3 May. As shown in panels (c)--(e), due to the mutual clockwise rotation between P2 and N2, 
P1 and N2 exhibited a collisional shearing motion \citep{2019ApJ...871...67C}. Note that the AR type was $\beta \delta$ and $\beta \gamma \delta$ on 3 and 4 May, respectively.

\begin{figure}
    \begin{center}
    \includegraphics[scale=0.9,bb=0 0 482 414]{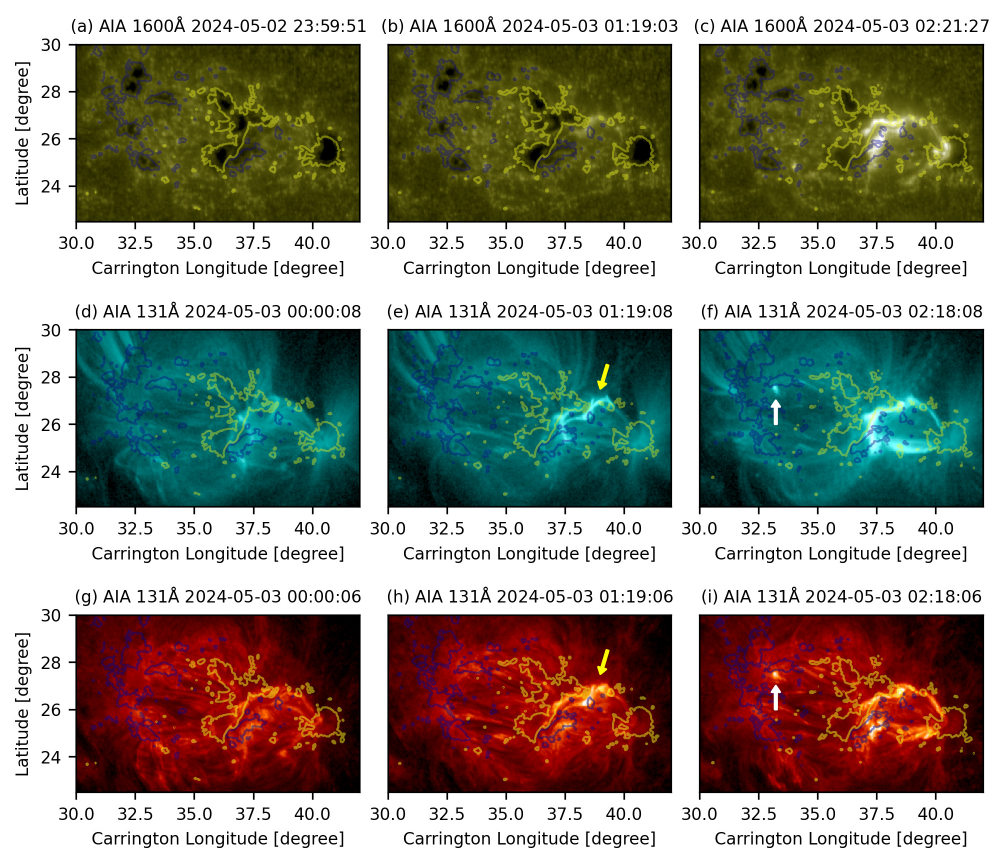}
    \caption{Time series of the multiwavelength AIA images before and during the X1.6 flare. The top, middle and bottom rows show the images of 1600 \AA, 131 \AA, and 304 \AA~channels, respectively. The columns correspond to the time. The line contours represent $B_{r}=600~\mathrm{G}$ (yellow) and $-600~\mathrm{G}$ (blue). The AIA images are reprojected onto the SHARP CEA coordinates The yellow and white arrows point to a transient brightening and a remote brightening, respectively.}
    \label{fig:aia_sharp}
    \end{center}
\end{figure}

Figure \ref{fig:aia_sharp} shows the time series of multiwavelength EUV images observed by AIA.
The AIA images are reprojected onto to the same CEA coordinate system as the SHARP magnetogram.
As shown in panels (e) and (h), there was a transient brightening prior to the flare from 01:17 UT to 01:24 UT at the northern part of the polarity inversion line (PIL) between P1 and N2. During the flare, in panels (c), (f) and (i), the bright ribbon structures appeared above the PIL between P1 and N2, and extended to N2. We also find a remote brighetening in the N1 region. The semi-circular flare ribbon kernel and the presence of the remote brightening are suggestive of a fan-spine magnetic configuration with a null point \citep{2012ApJ...757..149S,2013ApJ...771L..30J}.

\section{Methods}\label{sec:method}
\subsection{Data-driven model}\label{sec:data_driven_model}
We performed observational data-driven MHD simulations adopting
a velocity-driven method \citep{2021ApJ...909..155K,2024ApJ...974..168K}.
The simulation domain was constructed in a three-dimensional Cartesian coordinate system in which the $xy$-plane is a horizontal plane parallel to the photosphere, and the $z$-direction represents height.
The computational domain extends $0<x<223.6~\mathrm{Mm}$, $0<y<182.2~\mathrm{Mm}$, and $-3.376~\mathrm{Mm}<z<139.2~\mathrm{Mm}$
resolved by grid spacing sizes $\Delta x=0.844~\mathrm{Mm}$, $\Delta y=0.688~\mathrm{Mm}$, and $\Delta z=0.844~\mathrm{Mm}$,
corresponding to the grid numbers $N_{x}=N_{y}=N_{z}=256$, respectively.
In the domain above $z > 0$, the MHD equations as follows were numerically solved:
\begin{linenomath}
 \begin{equation}\label{eq:mass}
  \frac{\partial \rho }{\partial t} 
  + \nabla \cdot \left(\rho \boldsymbol{v} 
  \right)=0,
 \end{equation}
\end{linenomath}
\begin{linenomath}
 \begin{equation}\label{eq:mom}
  \frac{\left( \partial \rho \boldsymbol{v}\right)}{\partial t} 
  + \nabla \cdot \left(\rho \boldsymbol{v} \boldsymbol{v} 
  + p \boldsymbol{I} 
  - \frac{\boldsymbol{B}\boldsymbol{B}}{4\pi }
  + \frac{B^{2}}{8\pi }\boldsymbol{I}
  \right)=0,
 \end{equation}
\end{linenomath}
\begin{linenomath}
 \begin{equation}\label{eq:ene}
  \frac{\partial}{\partial t}\left(
  e_{\mathrm{th}}+\frac{1}{2}\rho v^{2}+\frac{B^{2}}{8\pi}
  \right)
  +\nabla \cdot \left[
  \left(e_{\mathrm{th}}+p+\frac{1}{2}\rho v^{2}\right)\boldsymbol{v}
  +\frac{1}{4\pi }\boldsymbol{E}\times \boldsymbol{B}
  \right]=0,
 \end{equation}
\end{linenomath}
\begin{linenomath}
 \begin{equation}\label{eq:iene}
 e_{\mathrm{th}}=\frac{p}{\gamma -1},
 \end{equation}
\end{linenomath}
\begin{linenomath}
 \begin{equation}\label{eq:induction}
 \frac{\partial \boldsymbol{B}}{\partial t}=-\nabla \times \boldsymbol{E},
 \end{equation}
\end{linenomath}
\begin{linenomath}
 \begin{equation}\label{eq:ele}
 \boldsymbol{E}=-\boldsymbol{v}\times \boldsymbol{B}
 +\eta \boldsymbol{J},
 \end{equation}
\end{linenomath}
\begin{linenomath}
 \begin{equation}\label{eq:current}
 \boldsymbol{J}=\frac{1}{4\pi }\nabla \times \boldsymbol{B},
 \end{equation}
\end{linenomath}
where $t$, $\rho $, $p$, $e_{\mathrm{th}}$, $\boldsymbol{v}$, $\boldsymbol{B}$, $\eta $, $\boldsymbol{J}$, and $\boldsymbol{I}$ represent time, mass density, gas pressure, internal energy, velocity field, magnetic field, resistivity, current density and unit tensor, respectively.
We applied an anomalous resistivity model as follows:
\begin{linenomath}
 \begin{equation}\label{eq:anomalous0}
 \eta = 0~~\mathrm{for}~~J<J_{\mathrm{th}},
 \end{equation}
\end{linenomath}
\begin{linenomath}
 \begin{equation}\label{eq:anomalous1}
 \eta = \min \left[\eta _{0}\left( J/J_{\mathrm{th}}-1\right)^{2},
 \eta _{\mathrm{max}}
 \right]~~\mathrm{for}~~J \ge J_{\mathrm{th}}
 \end{equation}
\end{linenomath}
where $J_{\mathrm{th}}=8\times 10^{-9}~\mathrm{G~cm^{-1}}$, 
$\eta _{0}=2\times 10^{13}~\mathrm{cm^{2}~s^{-1}}$, and $\eta _{\mathrm{max}}=2\times 10^{13}~\mathrm{cm^{2}~s^{-1}}$.
We assumed that the gravitational force is negligible as it is much smaller than the magnetic Lorentz force, and it does not significantly affect the flare dynamics.
These equations are numerically solved by the four-step Runge-Kutta method \citep{2005A&A...429..335V,2017AIAAJ..55.1487J} and a fourth-order central finite difference scheme with an artificial viscosity \citep{2014ApJ...789..132R}. 
To reduce the numerical errors caused by non-zero $\nabla \cdot \bm{B}$, we applied the hyperbolic divergence cleaning method \citep{2002JCoPh.175..645D}.

The 5 grids below $z \leq 0$ were divided into a driving layer (the upper 3 grids) and ghost cells (the lower 2 grids). In the driving layer, only Equations ($\ref{eq:induction}$) -- ($\ref{eq:current}$) were solved using the photospheric velocity $\boldsymbol{v}_{\mathrm{pho}}$ derived from the observed magnetograms. Note that, to compute the derivatives in $z$-direction, the coronal variables in $z>0$ were also partially involved.
$\boldsymbol{v}_{\mathrm{pho}}$ is computed as
\begin{linenomath}
 \begin{equation}\label{eq:exb}
    \boldsymbol{v}_{\mathrm{pho}}
    =\frac{\boldsymbol{E}^{I}\times \boldsymbol{B}_{\mathrm{sim}}}{B_{\mathrm{sim}}^{2}}   
 \end{equation}
\end{linenomath}
where $\boldsymbol{B}_{\mathrm{sim}}$ represents the simulated magnetic field, and $\boldsymbol{E}^{I}$ is the inverted electric field.
The $\boldsymbol{E}^{I}$ was obtained by inversely solving
\begin{linenomath}
\begin{equation}\label{eq:inversion}
    \frac{\boldsymbol{B}_{\mathrm{obs}}(t_{\mathrm{obs}}) 
    - \boldsymbol{B}_{\mathrm{sim}}(t)}
    {t_{\mathrm{obs}}-t} = -\nabla \times \boldsymbol{E}^{I},
    \end{equation}
\end{linenomath}
where $\boldsymbol{B}_{\mathrm{obs}}(t_{\mathrm{obs}})$ represents the observed magnetic field at the observation time $t_{\mathrm{obs}}$ of SDO/HMI, $\boldsymbol{B}_{\mathrm{sim}}(t)$ denotes the simulated magnetic field at the simulation time $t$, and $t_{\mathrm{obs}}$ is the closest time to $t$ satisfying $t_{\mathrm{obs}}>t$. In the simulations of this study, $\boldsymbol{E}^{I}$ was updated every $2~\mathrm{min}$ within the observational cadence of 12 min for SDO/HMI. Note that, since $\boldsymbol{B}_{\mathrm{sim}}$ is updated every time step, $\boldsymbol{v}_{\mathrm{pho}}$ is also updated every time step by Eq. \ref{eq:exb}. To solve Equation (\ref{eq:inversion}) inversely, we applied the poloidal-toroidal decomposition with a gauge to satisfy $\boldsymbol{E}\cdot \boldsymbol{B}=0$ \citep{2010ApJ...715..242F}. 
Thus, the magnetic field in the $z<0$ domain was forced to reproduce the observed photospheric magnetic field. Note that, as explained above, the observed magnetograms were not directly imposed as boundary condition. Instead, the observed magnetic field was reproduced through an assimilation 
between the observation and the simulation.
In the ghost cells, a free boundary condition was applied to density, pressure, and velocity. These quantities in the ghost cell were obtained by a simple zero-order extrapolation using the nearest grid points in the computational domain. The magnetic field was computed by solving the induction equation assuming the $z$-derivatives of $E_{x}$ and $E_{y}$ are zero. Since the induction equation for $B_{z}$ does not include $z$-derivative, the evolution of $B_{z}$ can be computed consistently. In contrast, the induction equations for $B_{x}$ and $B_{y}$ include $z$-derivatives, and therefore their evolution could not be fully reproduced. This is a limitation of data-driven simulations using the observed magnetograms, which do not provide information at multiple heights. To approximate an open boundary condition at the top boundary, a free boundary condition was applied to density, pressure, velocity, and the vertical component of the magnetic field. As the simplest way to maintain $\nabla \cdot \boldsymbol{B}=0,$ the horizontal components of the magnetic field were fixed to zero. For the side boundaries, periodic boundary conditions were applied to all variables because FFT was used in the electric field inversion. Since the top and the side bounadries were far from the reconnection sites, this boundary treatment should not significantly affect the main results on the triggering processes of the flare.

\subsection{Observational data and initial condition}
The vector magnetic field data from SDO/HMI SHARP for AR 13663 during 00:00 UT -- 05:00 UT on 3 May with a time cadence of 12 min were used for $\boldsymbol{B}_{\mathrm{obs}}$.
The simulation time $t=0$ corresponds to 00:00 UT on 3 May 2024.
The original grid size of SHARP data corresponds to $0.36~\mathrm{Mm}$.
We rebinned the observational data to fit the grid spacing size in the simulations.

To construct the initial three-dimensional force-free magnetic field, we performed a simulation in which potential field was gradually converted to the magnetic field at the first snapshot (00:00 UT). 
For the magnetic evolution, the same data-driven approach described in Section \ref{sec:data_driven_model} was applied between the potential field and the magnetic field at the first snapshot, while the plasma density and gas pressure were fixed to typical coronal values of $\rho = 1.67~\mathrm{g~cm^{-15}}$ and $p=0.276~\mathrm{erg~cm^{-3}}$, corresponding to temperature of $10^{6}~\mathrm{K}$. To suppress the unrealistically fast Alfv\'{e}n speed,
the magnetic field strength was reduced by a factor of $100$ from the original values.
After $\boldsymbol{B}_{\mathrm{sim}}$ became sufficiently close to $\boldsymbol{B}_{\mathrm{obs}}$, we set $\boldsymbol{v}_{\mathrm{pho}}=0$ and allowed the magnetic field to relax to a force-free state.

\subsection{Case study for post-input evolution}
The aim of this study is to examine whether a data-driven simulation can reproduce the flare by the observational data prior to the actual flare onset time. To this end, we assumed that no observational data were available after a certain time $t=t_{\mathrm{final}}$, and modified $\boldsymbol{v}_{\mathrm{pho}}$ in $t>t_{\mathrm{final}}$.
We tested two types of $\boldsymbol{v}_{\mathrm{pho}}$ models: Model S and Model E.
In Model S, we set $\boldsymbol{v}_{\mathrm{pho}}=0$ in $t>t_{\mathrm{final}}$. 
This means that the evolution of the photospheric magnetic field was completely stopped at the final observation time $t=t_{\mathrm{final}}$, 
and only the coronal magnetic field can evolve.
In Model E, we applied
\begin{linenomath}
\begin{equation}
    \boldsymbol{v}_{\mathrm{pho}}(t) = \boldsymbol{v}_{\mathrm{pho}}(t_{\mathrm{final}})\exp \left[ -\frac{t-t_{\mathrm{final}}}{\tau} \right], \label{eq:vpho}
\end{equation}
\end{linenomath}
where $\tau = 12~\mathrm{min}$.
This formulation allows the photospheric magnetic field 
to evolve under the velocity field at $t=t_{\mathrm{final}}$
within a characteristic decay timescale of 12 min.
The decay timescale is a free parameter, and we selected this value on a trial basis.
We changed $t_{\mathrm{final}}=72,~84,~96,~108,$ and $120~\mathrm{min}$ for each model, and examined whether the flare was reproduced at the actual flare onset time around $t=132~\mathrm{min}$ (start time) or $t=142~\mathrm{min}$ (peak time).

Hereafter, the case using all time-series magnetograms between 00:00 UT and 05:00 UT is referred to as the typical case.
The cases of Model S with $t_{\mathrm{final}}=72,~84,~96, 108$ and $120~\mathrm{min}$ are denoted as S72, S84, S96, S108, and S120, respectively.
Likewise, the cases of Model E are denoted as E72, E84, E96, E108, and E120.

\section{Results}\label{sec:result}
\subsection{Typical case}
\begin{figure}
    \begin{center}
    \includegraphics[scale=1.0,bb=0 0 450 457]{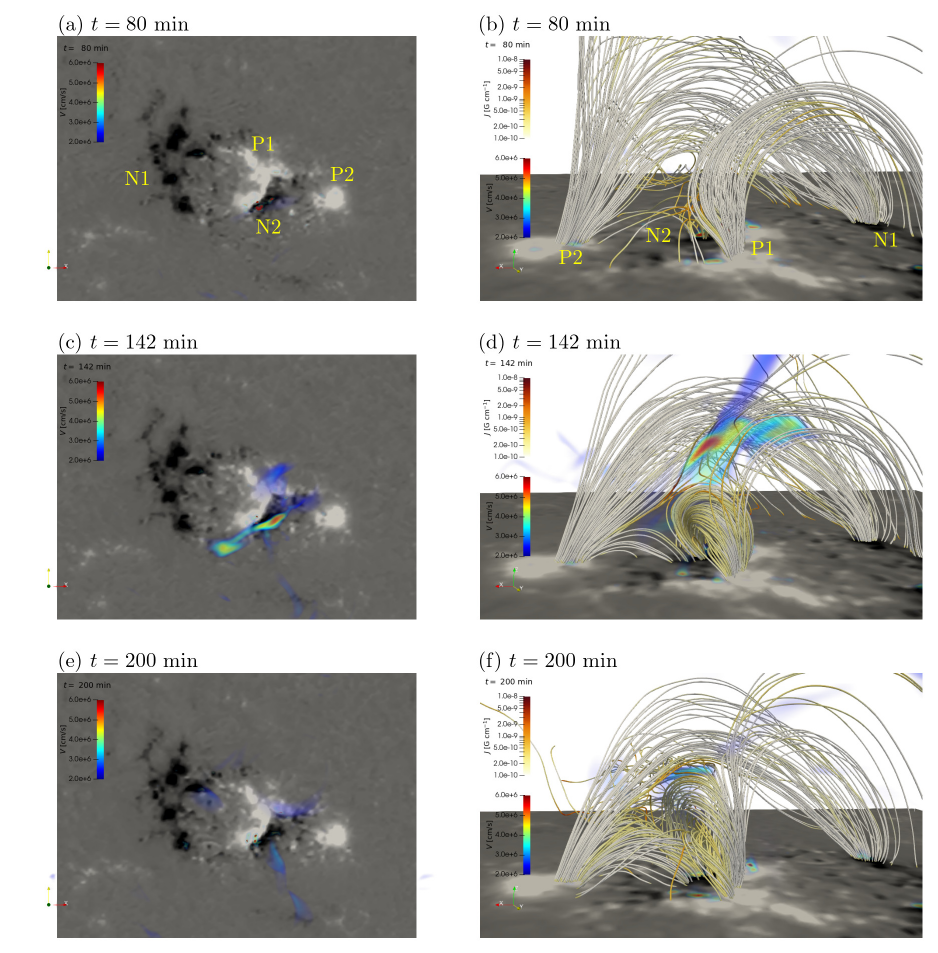}
    \caption{Temporal evolution of the MHD simulation for the typical case. Panels (a), (c), and (e) are rendered from bird's-eye perspective.
    Colors represent plasma speed $v$ (volume rendering of three-dimensional distribution), while the grayscale denotes the two-dimensional distribution of photospheric magnetic field component $B_{z}$. Panels (b), (d), and (f) show the same snapshots as (a), (c), (e), respectively, but rendered from a different perspective. The magnetic field lines are represented by the lines colored by the magnitude of current density. The movie version of this figure is provided as Additional file 1.}
    \label{fig:evol}
    \end{center}
\end{figure}

\begin{figure}
    \begin{center}
    \includegraphics[scale=1.0,bb=0 0 443 457]{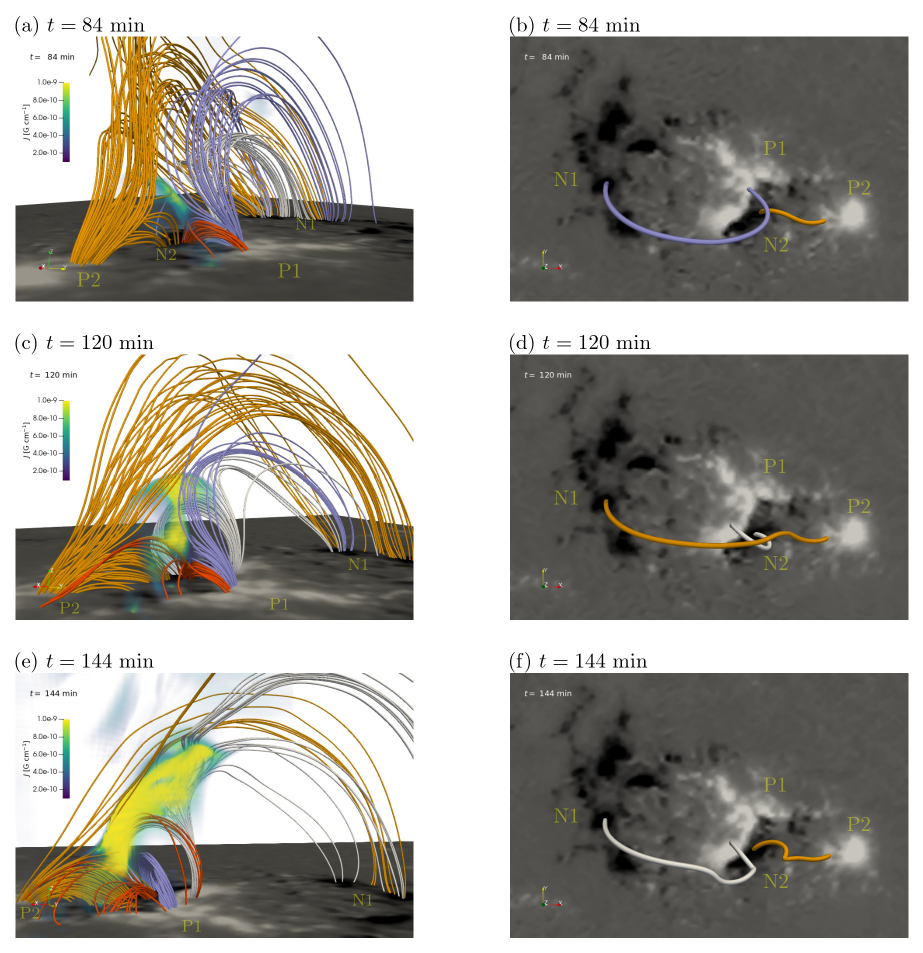}
    \caption{Temporal evolution of the current sheet and two-step reconnection process. Panels (a), (c) and (e) show volume renderings of the current density $J$. The field line colors represent the locations of the seed points used to trace the field lines. Panels (b), (d), and (f) show representative field lines illustrating the reconnection process. The times and the field line colors correspond to those in panels (a), (c), (e). The movie version of this figure is provided as Additional file 2.}
    \label{fig:rec}
    \end{center}
\end{figure}

\begin{figure}
    \begin{center}
    \includegraphics[scale=1.0,bb=0 0 319 379]{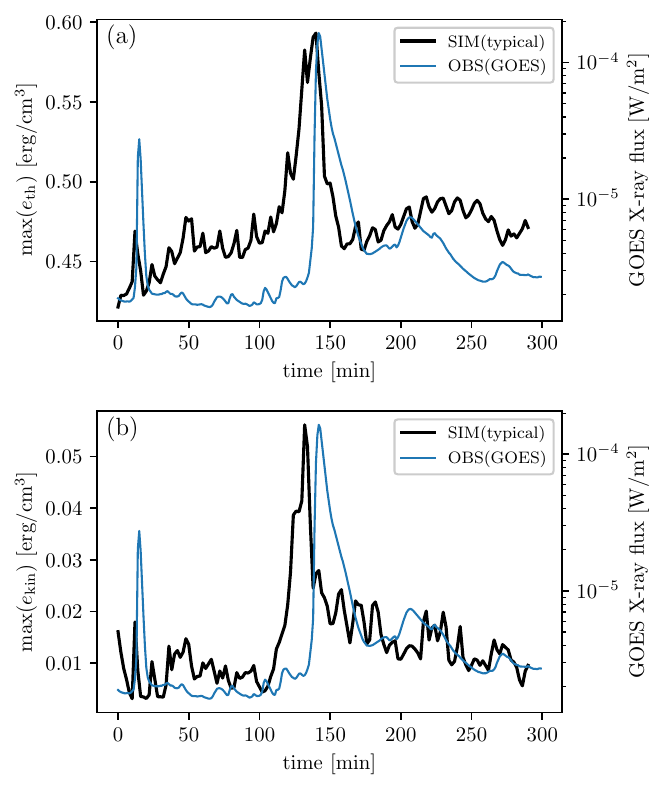}
    \caption{Panel (a): Temporal evolution of the maximum thermal energy density in the simulation and the X-ray flux observed by GOES. Panel (b): Same as panel (a), but for the maximum kinetic energy density.}
    \label{fig:comp_vs_goes}
    \end{center}
\end{figure}

\begin{figure}
    \begin{center}
    \includegraphics[scale=0.85,bb=0 0 447 387]{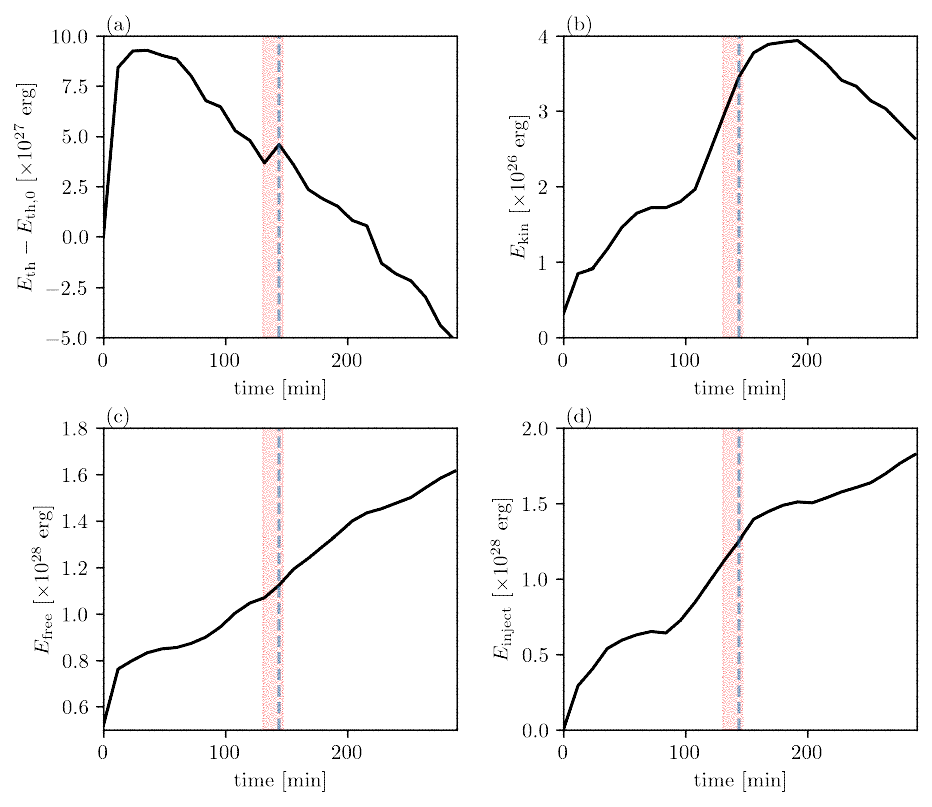}
    \caption{Panels (a), (b) and (c) show the temporal evolution of the thermal energy change $E_{\mathrm{th}}-E_{\mathrm{th},0}$, the kinetic energy $E_{\mathrm{kin}}$, and the free magnetic energy $E_{\mathrm{free}}$, respectively. Panel (d) shows the injected magnetic energy computed from the surface integration of the Poynting flux at $z=0.844~\mathrm{Mm}$. The red shaded area and the dashed line represent the flare duration and the peak time, respectively, defined by NOAA based on the GOES light curve.}
    \label{fig:ene_volume_typical}
    \end{center}
\end{figure}

\begin{figure}
    \begin{center}
    \includegraphics[scale=0.85,bb=0 0 548 427]{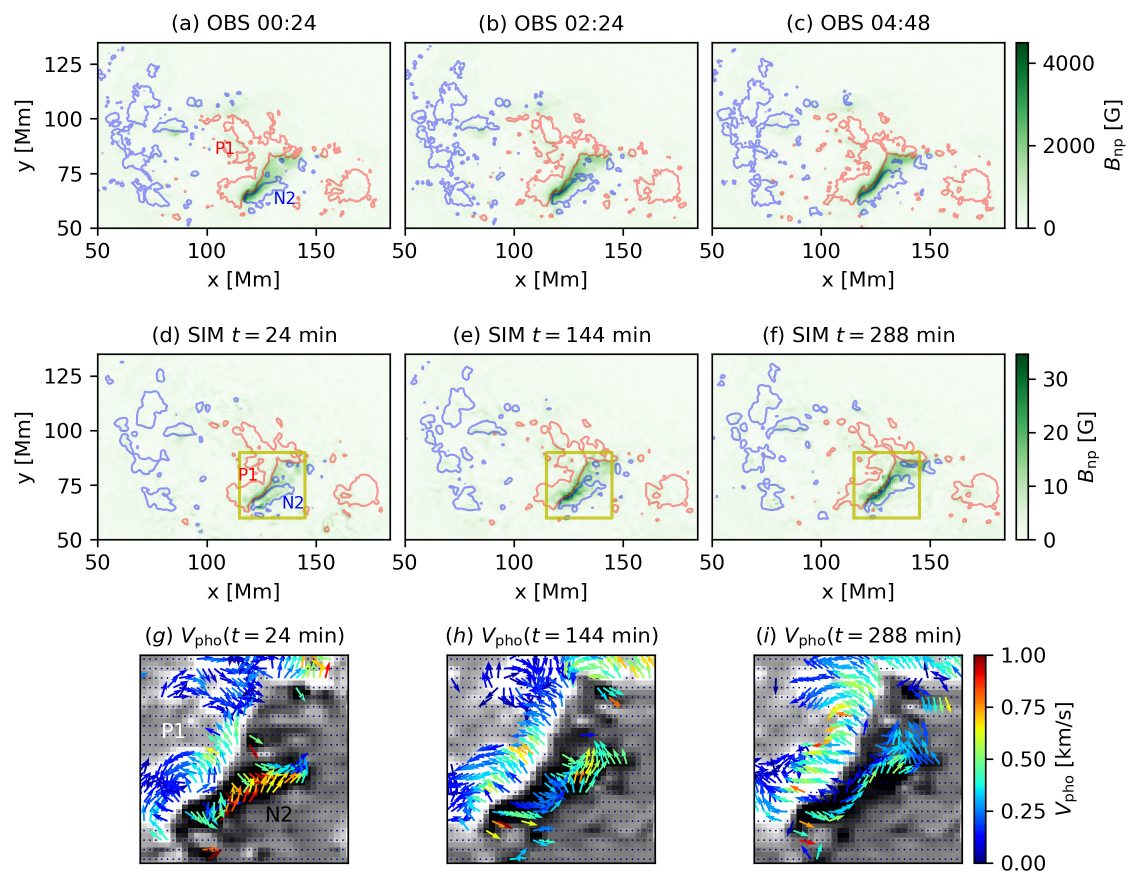}
    \caption{Panels (a)--(c) and (d)--(f) show the temporal evolution of $B_{\mathrm{np}}$ and $B_{z}$ in the observation and the simulation. The color denotes $B_{\mathrm{np}}$ . The blue and red line contours represent $B_{z}=-600~\mathrm{G}$ and $600~\mathrm{G}$, respectively. Panels (g)--(i) show the horizontal components of the photospheric velocity $\boldsymbol{v}_{\mathrm{pho}}$ within the yellow boxes in panels (d)--(f).}
    \label{fig:free_nonp_evol}
    \end{center}
\end{figure}

First, we present the simulation result using the all time-series magnetograms between 00:00 UT and 05:00 UT. Figure \ref{fig:evol} shows the temporal evolution of the magnetic structure and plasma velocity distribution.
As shown in panels (c) and (d), at $t=142~\mathrm{min}$ (the actual flare peak time), the plasma velocity over P1 and N2 patches rapidly increased.  Comparing panels (b), (d), and (f), the magnetic loops connecting P1 and N2 gradually expanded upward and reconnected with the overlying magnetic loops connecting P2 and N1. The rapid increase of the plasma velocity corresponds to the reconnection outflow.  
Figure \ref{fig:rec} depicts the evolution of current sheets and two-step reconnection.
  First, as shown in panel (a), a thin current sheet formed between the P1--N1 loops (purple) and the P2--N2 loops (lower orange lines). 
  Tether-cutting reconnection occurred at the current sheet above the northern part of the PIL between P1 and N2,
  creating the long magnetic loop P2--N1 (panels (b) and (d)). 
This reconnection would correspond to the observed transient brightening prior to the flare.
Then, as shown in panel (c), another current sheet formed at higher altitude between the overlying P2--N1 loops (orange) and the lower P1--N2 loops located in the southern part of the PIL (white). This current sheet has a broader distribution wrapping over the P1--N2 loops, suggesting the possible presence of null-point-like magnetic configuration. 
Finally, as shown in panels (d)--(f), the overlying P2--N1 loops and the lower P1--N2 loops reconnected, leading to the explosive energy release. This second reconnection corresponds to the flare.
Figure \ref{fig:comp_vs_goes} shows the temporal evolution of the maximum thermal energy density $\max \left(e_{\mathrm{th}}\right)$ and the maximum kinetic energy density $\max \left(e_{\mathrm{kin}}\right)$ in the $z>0$ region at each time, where
$e_{\mathrm{kin}}=0.5\rho |\boldsymbol{v}|^{2}$. They rapidly increased during the reconnection phase. Compared with the observed GOES X-ray flux profile, the peak times of $\max \left(e_{\mathrm{th}}\right)$ and $\max \left(e_{\mathrm{kin}}\right)$ are consistent with that of the actual flare. 
Figure \ref{fig:ene_volume_typical} shows the temporal evolution of the thermal energy change $E_{\mathrm{th}}-E_{\mathrm{th},0}$, the kinetic energy $E_{\mathrm{kin}}$, the free magnetic energy $E_{\mathrm{free}}$, and the injected magnetic energy $E_{\mathrm{inject}}$, respectively. 
Here, $E_{\mathrm{th},0}=6.16\times 10^{29}~\mathrm{erg}$ is the initial thermal energy.
The integration volume is 
$48~\mathrm{Mm}<x<185~\mathrm{Mm}$, 
$53~\mathrm{Mm}<y<142~\mathrm{Mm}$, 
and 
$0.844~\mathrm{Mm}\leq z<139.2~\mathrm{Mm}$. 
The injected magnetic energy $E_{\mathrm{inject}}$ is computed as the cumulative net Poynting flux as follows,
\begin{equation}
E_{\mathrm{inject}}(t)=\int _{0} ^{t} S_{z}(t^{\prime })dt^{\prime },
\end{equation}
where the $S_{z}$ is the net Poynting flux at $z=0.844~\mathrm{Mm}$ (1 grid above the driving layer)
defined as,
\begin{equation}
S_{z}=\int _{z=0.844~\mathrm{Mm}} \frac{1}{4\pi }(\boldsymbol{E}\times \boldsymbol{B})\cdot d\boldsymbol{S}.
\end{equation}
Note that the time integration for $E_{\mathrm{inject}}(t)$ was performed as a post-processing with a data output cadence of $2~\mathrm{min}$, whereas the simulation time step was of order of $1\times 10^{-2}~\mathrm{s}$. Therefore, the resulting values may contain numerical uncertainties. 
The boundary driving by the photospheric velocity $v_{\mathrm{pho}}$ leads to coronal heating and drives steady outflow from the corona \citep{2014MNRAS.440..971M,2019ApJ...880L...2S}. 
After $v_{\mathrm{pho}}$ was switched on at $t=0$, the thermal energy rapidly increased due to coronal heating. The subsequent gradual decrease in the thermal energy was due to mass loss by the steady outflows.
From the increment of the peaks in panels (a) and (b), the released thermal energy and the kinetic energy were approximately $1 \times 10^{27}~\mathrm{erg}$ and $1 \times 10^{26}~\mathrm{erg}$, respectively. Since the field strength $B$ in the simulation was reduced by a factor of 100 times from the observed value (see Section 3.2), and the magnetic energy is proportional to $B^{2}$, the released energy is estimated to be the order of $10^{31}~\mathrm{erg}$. This value is comparable to the typical released energy of the X1.6 flare. The free magnetic energy and the injected magnetic energy are comparable to each other, and are approximately one order of magnitude larger than the released energy. Because flux emergence continued throughout 3 May, the free magnetic energy and the injected energy monotonically increased during the simulation time. Because the energy injection dominated over the energy release, the decrease in the free magnetic energy associated with the flare was not clearly visible in panel (c). Note that the ratio of free magnetic energy compared to the total magnetic energy was $14~\%$ at the initial state, and it also monotonically increased up to $24~\%$.
Figure \ref{fig:free_nonp_evol} shows the temporal evolution of the non-potential magnetic field, defined as
\begin{equation}
B_{\mathrm{np}}=|\boldsymbol{B}-\boldsymbol{B}_{\mathrm{pot}}|,
\end{equation}
computed from the observed magnetogram (in the original spatial resolution) and from the simulated magnetic field in the driving layer. As shown in panels (g)--(i), the northern part of N2 continuously approached P1 and eventually collided with it. 
This converging motion led to the extension of the strong $B_{\mathrm{np}}$ region toward the northern part of the PIL, and it also triggered the first tether-cutting reconnection. The non-potential field may also be enhanced by the relative shear motion. 
The non-potential field was weaker in the northern part of the PIL and stronger in the southern part, suggesting the free magnetic energy was preferentially stored in the southern region. Therefore, the first tether-cutting reconnection did not cause explosive energy release, 
whereas the second breakout reconnection released a large amount of energy. The spatial distribution of $B_{\mathrm{np}}$ in the observation and the simulation are in good agreement.
However, even if the $B_{\mathrm{np}}$ is scaled up by the factor of 100, its magnitude reaches only 60--75$\%$ of the observed values.
This discrepancy may be due to the limitation of the velocity inversion method as well as the lower spatial resolution of the simulation compared with the observation. Therefore, the released energy estimated from the simulation would be a lower bound of the actual flare energy.

Thus, we show that the data-driven simulation using full time-series magnetograms captures the key processes for the onset of the X1.6 flare.

\subsection{Case study of post-input evolution}
\begin{figure}
    \begin{center}
    \includegraphics[scale=1.0,bb=0 0 282 379]{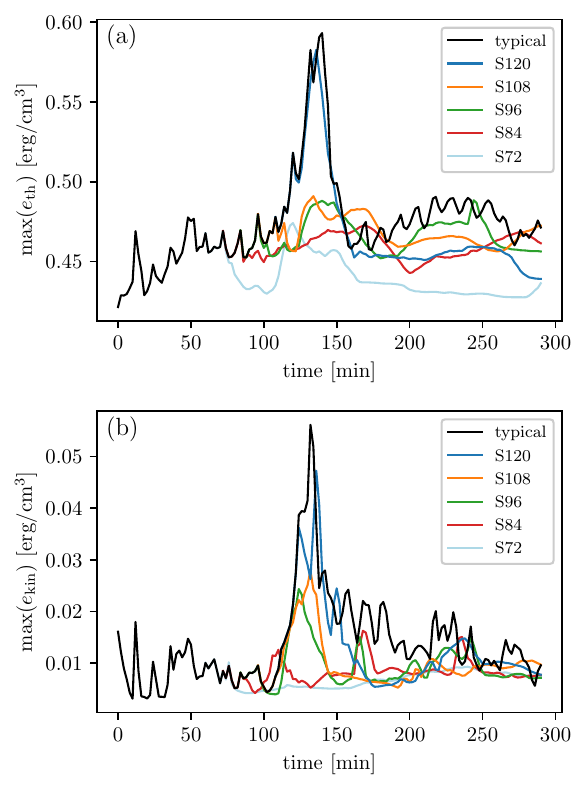}
    \caption{Panel (a): Temporal evolution of the maximum thermal energy density in each case for Model S. Panel (b): Same as panel (a) but for the maximum kinetic energy density.}
    \label{fig:comp_S}
    \end{center}
\end{figure}

Figure \ref{fig:comp_S} shows the temporal evolution of $\max (e_{\mathrm{th}})$ and $\max (e_{\mathrm{kin}})$ for each Model S case. S120 had a similar profile to the typical case. S108 and S96 show a clear enhancement in $\max (e_{\mathrm{kin}})$ around the actual flare onset time, whereas the enhancement in $\max (e_{\mathrm{th}})$ is less apparent. S84 and S72 do not show any clear enhancement in either $\max (e_{\mathrm{th}})$ or $\max (e_{\mathrm{kin}})$. Since the actual flare start and peak times are $t=132~\mathrm{min}$ and $t=142~\mathrm{min}$, respectively, the prediction lead time could be within approximately 1 hour.

\begin{figure}
    \begin{center}
    \includegraphics[scale=1.0,bb=0 0 282 379]{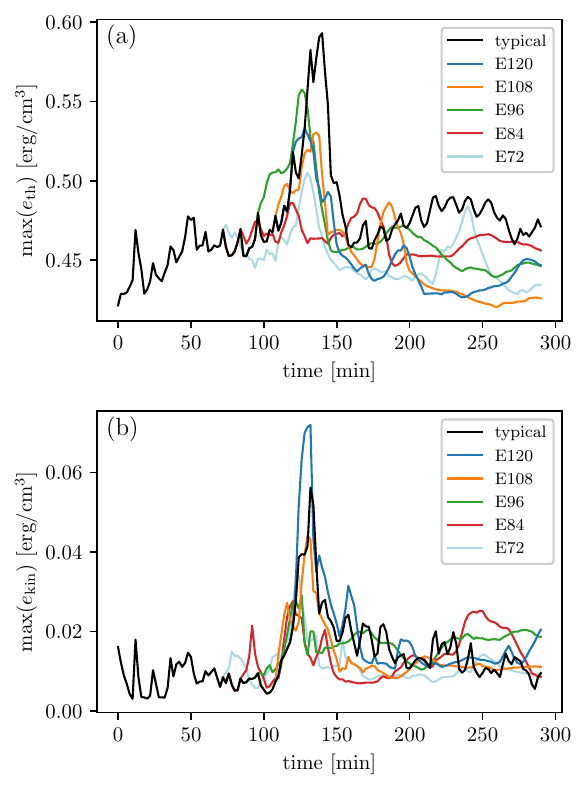}
    \caption{Same as Figure \ref{fig:comp_S} but for Model E.}
    \label{fig:comp_E}
    \end{center}
\end{figure}

Figure \ref{fig:comp_E} shows the temporal evolution of $\max (e_{\mathrm{th}})$ and $\max (e_{\mathrm{kin}})$ for each Model E case. All cases have peaks in either $\max (e_{\mathrm{th}})$ or $\max (e_{\mathrm{kin}})$ around the actual flare peak time. 
However, the magnitudes of these peaks vary somewhat randomly. The maximum energy densities in the early-stopping cases are not necessarily smaller, and vice versa.

\begin{figure}
    \begin{center}
    \includegraphics[scale=0.9,bb=0 0 444 387]{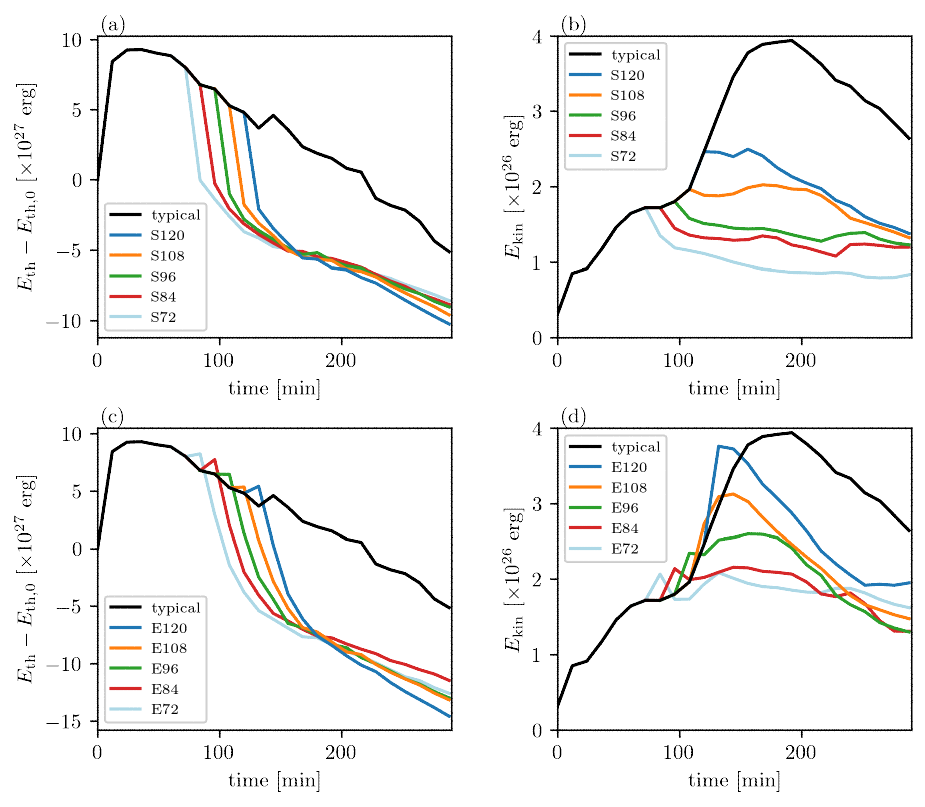}
    \caption{Panels (a) and (b) show the temporal evolution of the thermal energy change $E_{\mathrm{th}}-E_{\mathrm{th},0}$ and the kinetic energy $E_{\mathrm{kin}}$ in each case for Model S, respectively.  Panels (c) and (d) show the same quantities for Model E.}
    \label{fig:ene_volume_comp}
    \end{center}
\end{figure}

Figure \ref{fig:ene_volume_comp} shows the temporal evolution of $E_{\mathrm{th}}-E_{\mathrm{th},0}$ and $E_{\mathrm{kin}}$ for each case. 
The thermal energy dropped significantly after the boundary driving was stopped in both Model S and E cases. This is because the steady coronal heating sustained by the energy supply from the bottom boundary was stopped or decreased.
Overwhelmed by this entire decrease in thermal energy, the local enhancement by the flare reconnection was not clearly visible in Model S and E cases, unlike the typical case.
In Model S cases, the enhancement of kinetic energy was not clearly seen either. In contrast, in Model E cases, the enhancement of kinetic energy was more evident, and there was a systematic trend that the enhancement became smaller as the time of the boundary stopping was earlier.

\begin{figure}
    \begin{center}
    \includegraphics[scale=0.9,bb=0 0 431 690]{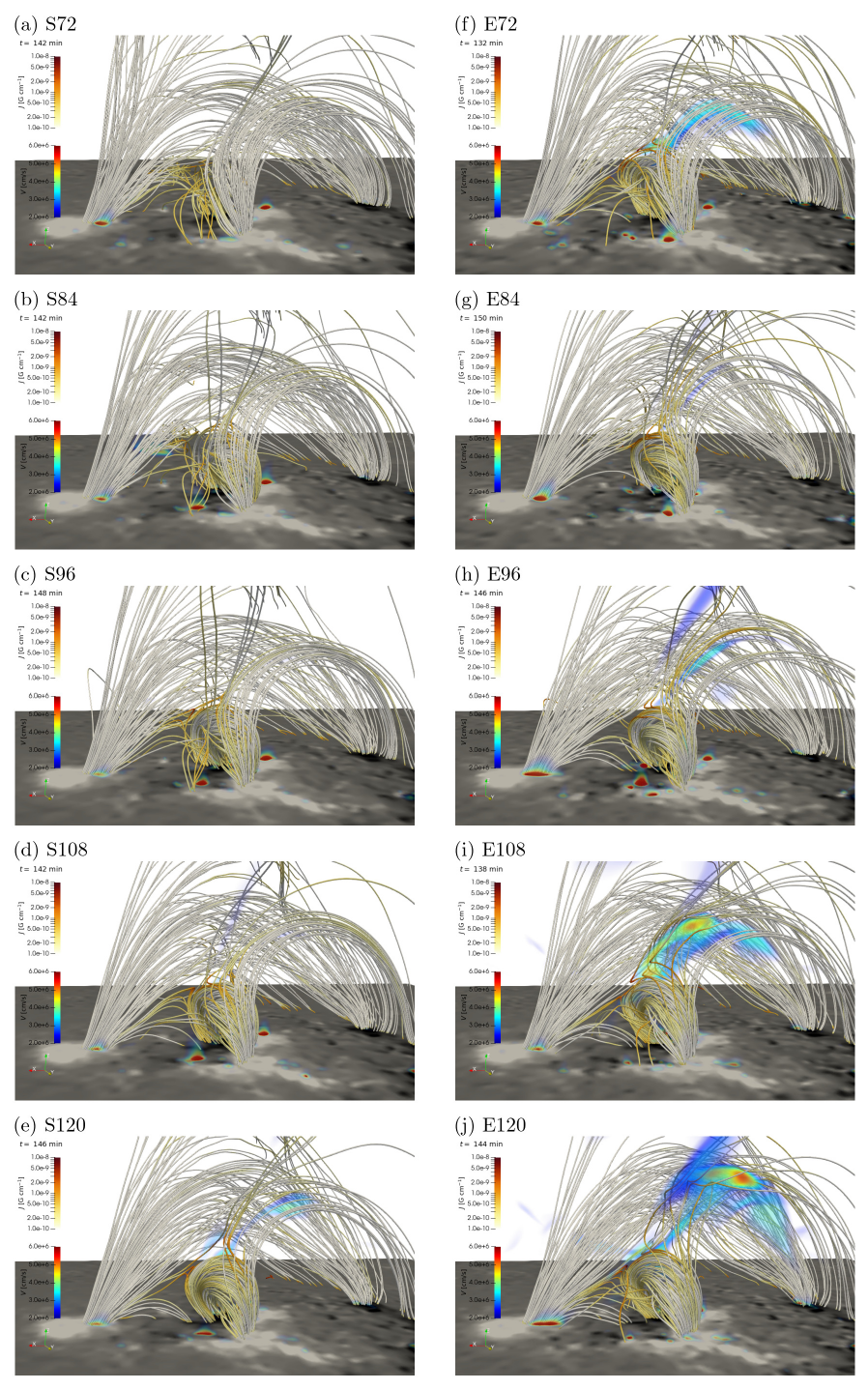}
    \caption{Magnetic structure in each case at around the actual flare onset time.}
    \label{fig:reconnection_comp}
    \end{center}
\end{figure}

Figure \ref{fig:reconnection_comp} displays the magnetic structures around the peak time for each case. In Model E, reconnection occurred in all cases.
In Model S, reconnection occurred in S120, S108 and S96. In S84, reconnection occurred quite gradually and the reconnection outflow was not prominent. In S72, reconnection did not occur. 

\begin{figure}
    \begin{center}
    \includegraphics[scale=0.9,bb=0 0 439 387]{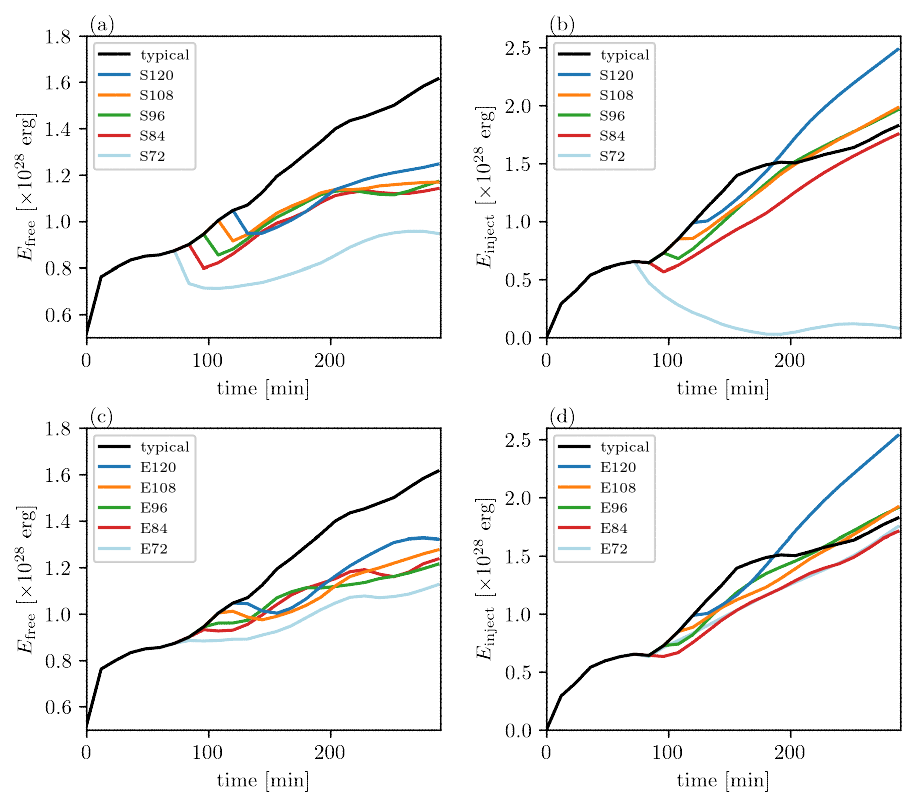}
    \caption{Panels (a) and (b) show the temporal evolution of the free magnetic energy $E_{\mathrm{free}}$ and the injected magnetic energy $E_{\mathrm{inject}}$ in each case for Model S, respectively. Panels (c) and (d) show the same quantities for Model E.}
    \label{fig:free_poy_comp}
    \end{center}
\end{figure}

Figure \ref{fig:free_poy_comp} shows the temporal evolution of $E_{\mathrm{free}}$ and $E_{\mathrm{inject}}$ in each case. The decrease in $E_{\mathrm{free}}$ during the flare was not clearly visible because the released thermal and kinetic energies were one or two orders of magnitude smaller than $E_{\mathrm{free}}$. Similar behavior was also reported in \citet{2016NatCo...711522J}.
In Model S cases, $E_{\mathrm{free}}$ and $E_{\mathrm{inject}}$ temporarily dropped after the boundary driving was stopped. This is likely due to numerical (artificial) diffusion acting on discontinuities created at the interface between the boundary layer and the coronal domain. In contrast, in Model E cases, the discontinuities were less likely to form because the photospheric velocity was gradually damped.
In S72, $E_{\mathrm{inject}}$ decreased to nearly zero after the boundary driving was stopped. In contrast, the other cases exhibited an increase in $E_{\mathrm{inject}}$.
Here, $E_{\mathrm{inject}}$ is defined at the surface one grid above the driving layer. Therefore, it is computed by the coronal velocity rather than $v_{\mathrm{pho}}$. Even when $v_{\mathrm{pho}}=0$, a finite value of $E_{\mathrm{inject}}$ can arise if coronal flows are present.
The decrease in $E_{\mathrm{inject}}$ indicates that the net Poynting flux was reversed after the boundary driving was stopped. In contrast, the increase in $E_{\mathrm{inject}}$ indicates that the energy flux was maintained by the coronal flows in the $z>0$ region. We assume that these flows are associated with the first tether-cutting reconnection based on the timing. Thus, we interpret that S84--S120 cases have already entered an irreversible self-driven state toward energy release. These results suggest a transition from the boundary-driven phase to a self-driven phase between S72 and S84.
In E72, the photospheric velocity was not abruptly stopped at $t=72~\mathrm{min}$, instead it was gradually damped with the time scale of $12~\mathrm{min}$. Hence, the system could enter the self-driven state.
Note that, in all cases, $E_{\mathrm{free}}$ showed an increasing trend even after the boundary driving was stopped. $E_{\mathrm{free}}$ was also evaluated based on the magnetic field at $z=0.84~\mathrm{Mm}$, where the magnetic field can still evolve temporally. The increasing trend resulted from both a slight increase in the total magnetic energy and a slight decrease in the potential magnetic energy. Although the variations in these two energies were individually small, their opposite trends resulted in a relatively larger variation in $E_{\mathrm{free}}$ difined as their difference. The exact mechanism for the small variation of each energy cannot be fully identified. 

\begin{figure}
    \begin{center}
    \includegraphics[scale=0.9,bb=0 0 443 457]{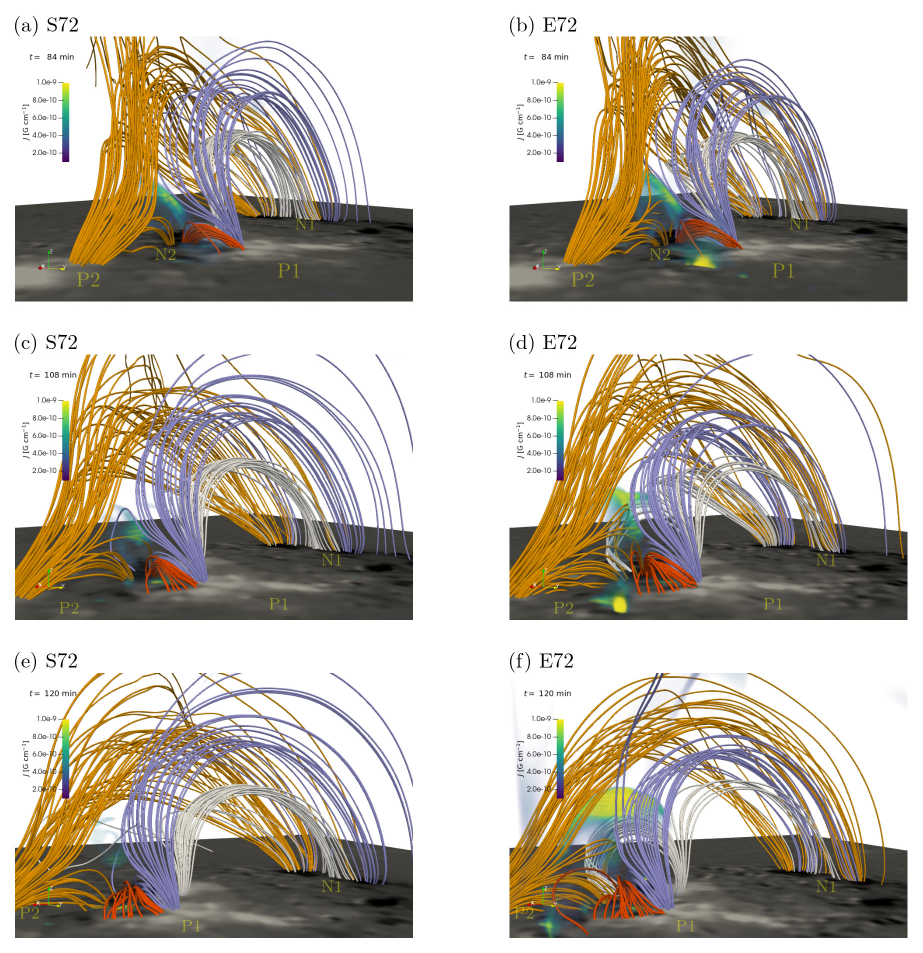}
    \caption{Temporal evolution of the current sheet in S72 and E72. 
    Panels (a), (c), and (e) show S72, while panels (b), (d), and (f) show E72. 
    The colors for the volume rendering and the lines represent the same as in Fig. \ref{fig:rec}.}
    \label{fig:rec_comp}
    \end{center}
\end{figure}

Figure \ref{fig:rec_comp} shows the temporal evolution of the current sheet in S72 and E72 around the time when the first reconnection occurred in the typical case. In E72, the first reconnection proceeded, leading to the formation of the current sheet for the second reconnection. In contrast, in S72, the current sheet faded and the first reconnection was not developed sufficiently. These results suggest that the first reconnection, which may correspond to the transient brightening in the observations, was a part of the irreversible, self-driven process.

\section{Discussion}\label{sec:discussion}
Our simulations suggest that two-step reconnection leads to the explosive energy release corresponding to the X1.6 flare. 
The first reconnection has features of tether-cutting \citep{2001ApJ...552..833M,2015ApJ...812...50J}. It occurs  between the P1--N1 loops and the P2--N2 loops in the northern part of the PIL, forming the long P2--N1 loops during $t=80-100~\mathrm{min}$. This first reconnection may correspond to the transient brightening observed prior to the flare. 
The second reconnection has features of breakout \citep{1999ApJ...510..485A,2008ApJ...683.1192L,2013ApJ...778..139S}. It occurs between the overlying P2--N1 loops and the lower P1--N2 loops in the southern part of the PIL during $t=100-150~\mathrm{min}$. 

Flares in the quadrupolar magnetic distribution have been investigated in several previous studies. 
\citet{2012ApJ...757..149S} analyzed an M2.2 flare in AR 11158, and found
that a flux rope was confined within the fan-spine magnetic field with a null point. They discussed that the eruption was triggered after the null point reconnection which reduced the confinement to the flux rope. A similar mechanism was reported for the X2.1 flare in AR 11283 by \citet{2013ApJ...771L..30J}, where the reduction of confinement after a null point reconnection led to the torus instability of the flux rope.
In contrast, \citet{2024ApJ...975..206F} and \citet{2025MNRAS.541..939D} claimed that the initiation of X2.2 flare in AR 11158 was the tether-cutting reconnection in a pre-flare current sheet. 
\cite{2015ApJ...812...50J} proposed multi-stage successive reconnection scenario for M7.3 flare in AR 12036, in which tether-cutting reconnection first created and erupted a sigmoid, followed by null point reconnection at higher altitudes. 
Our results resemble this multi-stage reconnection scenario, however; the driver of the second breakout reconnection was not a rising sigmoid or flux rope. In our simulations, the tether-cutting reconnection created the long twisted magnetic fields (P2--N1), which facilitated the breakout reconnection as overlying fields (see Fig. \ref{fig:rec} (b), (d) and (f)).
As mechanisms of flares and eruptions, MHD instabilities of a flux rope, such as kink instability, torus instability, and double-arc instability have also been proposed \citep{1979SoPh...64..303H,2006PhRvL..96y5002K,2017ApJ...843..101I}. 
However, the actual event analyses by \citet{2019ApJ...884...73D} reported that the criteria of the kink and torus instabilities were not necessarily satisfied prior to flares, highlighting the importance of reconnection as the initiation process. Our results are consistent with their view, in that the flare onset was reproduced primarily by reconnection. Note that the torus instability may play a crucial role for the subsequent CME.

Our results of post-input magnetic evolution suggest that, if the photospheric boundary was completely fixed at the state of the final observation time, the prediction lead time for the X1.6 flare would be roughly one hour. If the information of photospheric velocity field (or equivalently, the electric field) was incorporated, the lead time could be extended. We revealed the transition from the boundary-driven phase to a self-driven phase during $t=72-84~\mathrm{min}$. In S72, the boundary input was stopped before entering the self-driven phase, while in E72, the system could enter the self-driven phase because the velocity input continued after $t=72~\mathrm{min}$ instead of complete termination.
Although this study did not identify a universal threshold for the transition to the self-driving phase, in this specific event analysis, the results suggest that the relative location between the tether-cutting and breakout reconnection sites might affect the threshold. If the two reconnection sites are closer to each other, weaker tether-cutting reconnection can be sufficient to trigger the subsequent breakout reconnection, and vice versa. However, such triggering conditions are expected to strongly depend on the magnetic configuration.
\citet{2016NatCo...711522J} also investigated post-input evolution using data-driven simulations for a flare produced by AR 11283. In their study, when the boundary input was stopped several minutes before the flare onset, the flare was not reproduced. Depending on the properties of the active regions and the triggering mechanism of flare, such as combination of multi-step reconnection and MHD instabilities, the predictability of flares may vary. Further studies on various types of flares and active regions should be carried out.

There are two categories of data-driven models \citep{2020ApJ...890..103T}. 
One is the $\boldsymbol{B}$-driven model in which the observed magnetograms are directly used as the boundary condition for the magnetic field. The other is the $\boldsymbol{E}$-driven model in which the electric field is used to reproduce the temporal evolution of the photospheric 
magnetic field. The velocity-driven method applied in this study is categorized as an $\boldsymbol{E}$-driven model. 
In practical prediction, however, neither $\boldsymbol{B}$ nor $\boldsymbol{E}$ in the future can be obtained. Therefore, the accurate long-term prediction is inherently difficult. Nevertheless, the $\boldsymbol{E}$-driven model has the potential to extend the prediction lead time by integrating the induction equation using the electric field at the final observation time, thereby inferring the magnetic field near future. In this study, we used the velocity field model represented by Eq. (\ref{eq:vpho}) as a trial attempt. Further studies are necessary to find an appropriate photospheric velocity or electric field model in the post-input stage.

The quantitative reproduction of the stored and released energy remained challenging even in the typical case of our simulation. One possible reason is the lower spatial resolution in the simulations compared to the observations. Another possible reason is that the dependency of the initial condition. In the present study, the discrepancy in $B_{\mathrm{np}}$ between the observation and the simulation was relatively larger at the initial state, but it decreased as the simulation progressed.  This is because our data-driven method has an aspect of data-assimilation, allowing the system gradually adjust to the observational state during the temporal integration. If the initial state could be set much earlier than the flare onset, the reproduced energy might be more quantitatively consistent. However, the use of earlier vector magnetograms was limited by data quality. In the data on 2 May, spurious polarity fluctuations (i.e., the magnetic polarity in the low signal regions repeatedly flips between positive and negative in the consecutive snapshots) are present inside the AR, likely because the AR position was not sufficiently close to the disk center. Because data-driven simulations involve time series data, such a temporally-inconsistent artifacts can significantly degrade the accuracy and reliability of the results. Hence, we set the initial state at 00:00 UT on 3 May in the present study. A temporally consistent dataset, such as SuperSynthIA generated by deep learning techniques \citep{2024ApJ...970..168W}, would be helpful to address this limitation

\section{Conclusions}\label{sec:conclusion}
We revealed that the X1.6 flare on 3 May 2024 was triggered by two-step reconnection. The first tether-cutting reconnection due to converging motion between the opposite polarities (P1 and N2 in the northern area) created the overlying long loop (P2-N1). The second breakout reconnection occurred between the overlying long loop (P2-N1) and the inner short loop (P1-N2 in the southern area), leading to the explosive energy release.

We demonstrated that the reconnection responsible for the flare could be reproduced roughly one hour prior to the flare onset time, even if the magnetogram at the flare onset time was not available. We also showed that, when the information of photospheric velocity field (or electric field) is incorporated, the prediction lead time can be extended. These results demonstrate the potential of data-driven MHD simulations for operational flare prediction.

\section*{Declarations}
\section*{Availability of data and materials}
The datasets used and analyzed during the current study are available from the corresponding author upon reasonable request.

\section*{Competing interests}
The author declares that there are no competing interests.

\section*{Funding}
This work was supported by JSPS KAKENHI Grant Number 20K14519.

\section*{Authors' contributions}
The author carried out all simulations, analyses, and manuscript preparation.

\section*{Authors' information}
${}^{1}$ Faculty of Education, Niigata University, 8050 Ikarashi 2-no-cho, Nishi-ku, Niigata, 950-2181, Japan

${}^{2}$ Department of Electrical and Information Engineering, Graduate School of Science and Technology, Niigata University, 8050 Ikarashi 2-no-cho, Nishi-ku, Niigata, 950-2181, Japan

\acknowledgments{
We are grateful to the anonymous referees for their constructive and insightful comments. This work was supported by JSPS KAKENHI Grant Number JP20K14519. Numerical computations were conducted on a Cray XC50 supercomputer at the Centre for Computational Astrophysics (CfCA) of the National Astronomical Observatory of Japan. HMI and AIA are the instruments on the SDO, a mission for NASA's Living with a Star program. The GOES X-ray flux data were obtained from the National Centers for Environmental Information (NCEI), NOAA. This research used version 7.0.3 of the SunPy open source software package.
}

\providecommand{\url}[1]{{#1}}
\providecommand{\urlprefix}{URL }
\expandafter\ifx\csname urlstyle\endcsname\relax
  \providecommand{\doi}[1]{doi:\discretionary{}{}{}#1}\else
  \providecommand{\doi}{doi:\discretionary{}{}{}\begingroup
  \urlstyle{rm}\Url}\fi
\providecommand{\apj}{Astrophys~J~}
\providecommand{\apjl}{Astrophys~J~Lett}
\providecommand{\solphys}{Sol~Phys~}
\providecommand{\aap}{Astron~Astrophys}
\providecommand{\prl}{Phys~Rev~Lett~}
\providecommand{\lrsp}{Living~Rev~Sol~Phys~}
\providecommand{\mnras}{Mon~Not~R~Astron~Soc}


\end{document}